\documentclass[twocolumn,aps,pra,amsmath,amssymb,showpacs]{revtex4-1}
\usepackage[latin9]{inputenc}
\usepackage{color}
\usepackage{amsmath}
\usepackage{graphicx}
\usepackage[unicode=true,pdfusetitle,
 bookmarks=true,bookmarksnumbered=false,bookmarksopen=false,
 breaklinks=false,pdfborder={0 0 0},pdfborderstyle={},backref=false,colorlinks=true]
 {hyperref}
\hypersetup{
 colorlinks,linkcolor=red,citecolor=blue}
\usepackage{breakurl}
\usepackage{epsfig}
\usepackage{amsfonts}
\usepackage[english]{babel}



\begin{document}

\title{Quantum states preparation of an atomic ensemble via cavity-assisted
homodyne measurement}

\author{Yan-Lei Zhang$^{1,2}$ }

\author{Chuan-Sheng Yang$^{1,2}$ }

\author{Chang-Ling Zou$^{1,2}$ }
\email{clzou321@ustc.edu.cn}

\author{Tian Xia$^{2,3}$ }

\author{Guang-Can Guo$^{1,2}$ }

\author{Xu-Bo Zou$^{1,2}$ }
\email{xbz@ustc.edu.cn}

\affiliation{$^{1}$Key Laboratory of Quantum Information, University of Science
and Technology of China, CAS, Hefei, Anhui 230026, China}

\affiliation{$^{2}$CAS Center for Excellence in Quantum Information and Quantum
Physics, University of Science and Technology of China, Hefei, Anhui
230026, China}

\affiliation{$^{3}$Hefei National Laboratory for Physical Sciences at the Microscale,
University of Science and Technology of China, Hefei, 230026 China}

\date{\today}
\begin{abstract}
The quantum spin states of atomic ensemble are of special interesting
for both fundamental studies and precision measurement applications.
Here, we propose a scheme to prepare collective quantum states of
an atomic ensemble placed in an optical cavity via homodyne measurement
of probing light field. The effective interactions of atoms mediated
by photons are enhanced by the optical cavity, and the output probe
light could also be entangled with the collective spin states. By
selectively measuring the quadrature of output light, we can prepare
various quantum states, including superposition states of Dicke states
and Dicke squeezed states. It is also demonstrated that the fidelity
of prepared quantum state can be enhanced by repetitive homodyne detection
and using longer probe laser pulses. Our scheme is feasible for experimental
realization with current technologies, which may be used in future
study of quantum mechanics and quantum metrology.
\end{abstract}

\maketitle

\section{Introduction}

Large ensembles of atoms are good platforms for testing fundamental
physics \cite{hammerer2010quantum,safronova2018search} and practical
applications \cite{duan2001long,sangouard2011quantum}, such as, quantum
metrology \cite{cronin2009optics,hosten2016quantum}, quantum memory
\cite{nunn2008multimode}, atomic clocks \cite{borregaard2013efficient,ludlow2015optical,komar2014quantum}
and gravitational wave detectors \cite{khalili2018overcoming,graham2013new}.
For these applications, the preparation of quantum states of atomic
ensembles \cite{frowis2018macroscopic,hammerer2010quantum,pettersson2017light}
is essential. For example, the quantum superposition states \cite{cirac1998quantum,frowis2018macroscopic,mermin1990extreme,sarkar2018high}
and Dicke squeezed states (DSS) \cite{zhang2014quantum} are the two
typical quantum states, which show interesting quantum phenomena and
break classical limitation by utilizing the entanglement property
of collective spins \cite{kuzmich2000generation,toth2012multipartite,demkowicz2014using}.
The quantum superpositions of collective spins are allowed by quantum
mechanics, including Greenberger-Horne-Zeilinger (GHZ) \cite{bollinger1996optimal,zou2003conditional}
and W states \cite{yu2007robust,zheng2006splitting}, which can be
applied for quantum information processing. The spin squeezing states
\cite{kitagawa1993squeezed} and DSS \cite{zhang2014quantum} are
many-particle entangled states, allowing the quantum metrology beyond
the standard quantum limit \cite{ma2011quantum,wineland1992spin,lucke2014detecting}.
Therefore, the preparation of these quantum states is of great importance
and has attracted considerable attention recently.

Many schemes \cite{masson2017cavity,engelsen2017bell,ma2011quantum,wang2017two}
have been proposed to prepare quantum states of atomic ensemble spins
in recent years. Spin squeezing and entanglement of collective spins
have been achieved by directly transferring squeezing from squeezed
light to an atomic ensemble \cite{fleischhauer2002stationary}. In
analogy with nonlinear optics, quantum states of collective spins
can be generated via collisional interactions in Bose-Einstein condensate
\cite{esteve2008squeezing,helmerson2001creating,huang2015two}. Alternatively,
cavity-based schemes \cite{zhang2015detuning,zhang2015phonon} have
also been proposed to realize highly squeezed states and other quantum
states, where the light-matter interaction is enhanced by placing
the atoms in an optical cavity \cite{schleier2010squeezing,leroux2010implementation,dalla2013dissipative}.
The effective interactions of collective atoms are induced by the
optical cavity, while the interaction types can not be chosen at will,
and ideal quantum states can not be prepared. It is also demonstrated
that the quantum nondemolition (QND) detection on the states of photons
coupled with the atomic ensemble could be utilized to produce quantum
states \cite{inoue2013unconditional}, including the entanglement
of two macroscopic atomic samples \cite{julsgaard2001experimental},
spin squeezing states \cite{leroux2012unitary,kuzmich2000generation},
and Dicke state \cite{vanderbruggen2011spin,stockton2004deterministic}.
However, a scheme that combines the advantages of cavity-induced nonlinear
couplings and the variety of QND detection has never been proposed
yet.

In this work, we propose an experimental feasible scheme to prepare
collective quantum states of an atomic ensemble by cavity assisted
homodyne measurements. For a cavity-atom ensemble system probed by
coherent resonant laser pulse, the output light is entangled with
the collective spin states of atomic ensemble in the cavity, which
induces linear, square, and high order nonlinear couplings of collective
spins. By time-domain homodyne measuring of the quadrature of output
light, the quantum superposition state and DSS can be created selectively.
For a given system, the state preparation can be optimized by carefully
choosing the measurement strength and pulse length. Moreover, the
state preparation can be enhanced by repetitive homodyne measurement.
Our scheme is feasible for experimental realization and can also be
generalized to generate other non-Gaussian states by optimal homodyne
measurement.

\begin{figure}[htbp]
\center \includegraphics[width=0.8\columnwidth]{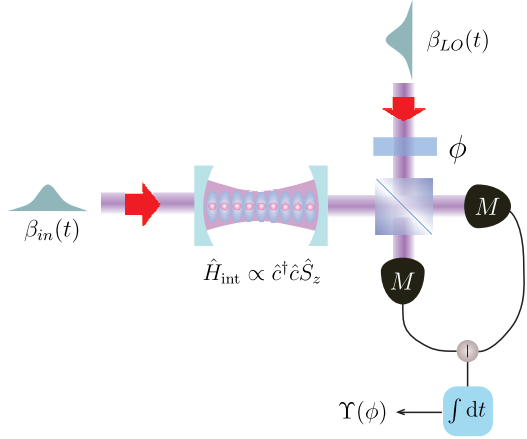}

\caption{(Color online) Schematic illustration of cavity-atom ensemble system
with an optical quadrature measurement performed via homodyne detection.
By choosing the phase $\phi$ of the local laser, we can realize the
time-domain homodyne detection $\Upsilon\left(\phi\right)$ of the
output light, which leads to non-Gaussian states of the collective
spins.}

\label{Fig1}
\end{figure}

\section{Model}

Figure$\,$\ref{Fig1} is a schematic of the cavity-atom ensemble
system setup. We use a laser pulse resonant with the cavity to probe
the system, which contains an ensemble of $N$ identical atoms with
two stable ground states and an excited state. By adiabatically eliminating
the excited state of atoms, the two ground levels of atom ensemble
are coupling to the optical cavity \cite{schleier2010squeezing,zhang2015detuning},
with the effective Hamiltonian as
\begin{equation}
H_{\mathrm{eff}}=\Omega c^{\dagger}cS_{z}-i\sqrt{2\kappa N_{p}}(c-c^{\dagger})\beta_{in}\left(t\right),\label{eq:Hamiltonian}
\end{equation}
where $c,~c^{\dagger}$ are the photon annihilation and creation operators
for the cavity mode, $2S_{z}=\sum_{i=1}^{N}(\left|\uparrow\right\rangle _{i}\left\langle \uparrow\right|_{i}-\left|\downarrow\right\rangle _{i}\left\langle \downarrow\right|_{i})$
is $z$ component of the collective spin operator, and the coefficient
$\Omega=2g^{2}/\left|\Delta\right|$ is the dispersive cavity and
atom population coupling strength \cite{schleier2010squeezing}, in
which $g$ is the coupling strength between the atom ground-excited
transitions and vacuum cavity field, and $\Delta$ is the detuning
between the cavity mode and atom transitions. The probe pulse has
a mean photon number $N_{p}$ with the normalized pulse envelope $\int dt\left|\beta_{in}(t)\right|^{2}=1$,
and $\kappa$ is the cavity amplitude decay rate.

We notice that the spin operator $S_{z}$ is a conserved quantity
in the system, since $\left[S_{z},H_{\mathrm{eff}}\right]=0$. Therefore,
by accumulating an extra phase in proportion to the spin operator
$S_{z}$ \cite{zhang2015detuning}, the cavity field evolves by following
the Heisenberg equation
\begin{equation}
\frac{d}{dt}c=\left(-\kappa-i\Omega S_{z}\right)c+\sqrt{2\kappa N_{p}}\beta_{in}\left(t\right).\label{eq:cMotion}
\end{equation}
 The formal solution \cite{milburn1983quantum} of cavity field for
$c\left(-\infty\right)=0$ is
\begin{equation}
c\left(t\right)=\sqrt{2N_{p}\kappa}\int_{-\infty}^{t}dt'\exp\left[-\left(\kappa+i\Omega S_{z}\right)\left(t-t'\right)\right]\beta_{in}(t').\label{eqC-2}
\end{equation}
As the term in proportion to the spin operator is assumed to be small,
the mean intracavity field can be approximated as \cite{vanner2011selective}
\begin{equation}
\left<c\left(t\right)\right>/\sqrt{N_{p}}\simeq\beta_{0}(t)-i\frac{\Omega}{\kappa}\beta_{1}(t)\left<S_{z}\right>-\frac{\Omega^{2}}{2\kappa^{2}}\beta_{2}(t)\left<S_{z}^{2}\right>,\label{eqC-1}
\end{equation}
where the dimensionless functions are
\begin{align}
\beta_{0}(t) & =\sqrt{2\kappa}\int_{-\infty}^{t}dt'\exp\left[-\kappa\left(t-t'\right)\right]\beta_{in}(t'),\\
\beta_{1}(t) & =\sqrt{2\kappa}\int_{-\infty}^{t}dt'\exp\left[-\kappa\left(t-t'\right)\right]\kappa\left(t-t'\right)\beta_{in}(t'),\\
\beta_{2}(t) & =\sqrt{2\kappa}\int_{-\infty}^{t}dt'\exp\left[-\kappa\left(t-t'\right)\right]\kappa^{2}\left(t-t'\right)^{2}\beta_{in}(t').
\end{align}

Since the cavity field contains information of spin operator $S_{z}$,
the spins can be projected to certain collective spin states by time-domain
homodyne detection of the output field $c_{out}=\sqrt{2\kappa}c$.
The homodyne measurement operator \cite{vanner2011selective} of the
output reads
\begin{equation}
\Upsilon\left(\phi\right)=\sqrt{2}\int dt\beta_{LO}(t)\Upsilon^{out}(\phi),
\end{equation}
where $\Upsilon^{out}(\phi)=\frac{1}{\sqrt{2}}\left(c_{out}^{\dagger}e^{i\phi}+c_{out}e^{-i\phi}\right)$
is the measured quadrature, and $\phi$, $\beta_{LO}$ are the phase
and amplitude of the local pulse, respectively, as shown in Fig.$\,$\ref{Fig1}.
The corresponding measurement operator \cite{vanner2011pulsed} reads
\begin{equation}
M=\left(\pi\right)^{-\frac{1}{4}}\exp\left[i\eta S_{z}-\left(\Upsilon+\chi_{x}S_{z}^{2}+\chi_{p}S_{z}\right)^{2}/2\right],\label{eq:Ry}
\end{equation}
by generalised linear measurement theory \cite{caves1987quantum}.
Here $\eta$ is the accumulated phase,
\begin{equation}
\chi_{x}=\sqrt{2\kappa}\int dt\frac{\Omega^{2}}{\kappa^{2}}\beta_{LO}(t)\beta_{2}(t)\cos(\phi),\label{eq:chix}
\end{equation}
 is the square-spin operator measurement strength, and
\begin{equation}
\chi_{p}=2\sqrt{2\kappa}\int dt\frac{\Omega}{\kappa}\beta_{LO}(t)\beta_{1}(t)\sin(\phi),\label{eq:chip}
\end{equation}
 is the linear spin operator measurement strength, all of which are
determined by the local pulse \cite{vanner2011selective}. As the
term from $\beta_{0}$ does not contain any information of the spin
operator, we have neglected this contribution above. Since the quadratures
of intracavity field carry information of the spin operators $S_{z}$
and $S_{z}^{2}$ respectively, the time-domain homodyne measuring
of the quadrature can generate quantum states of the collective spins,
as indicated by the measurement operator. By properly choosing the
phase $\phi$ and shaping the local pulse $\beta_{LO}$, we can selectively
imprint the actions $S_{z}^{2}$ and $S_{z}$ on the atom ensemble
and generate quantum states of the atom ensemble.

\section{Quantum superposition of Dicke states}

When the phase of local oscillator is chosen as $\phi=0$, we measure
the X quadrature $\Upsilon^{out}\left(0\right)=X^{out}\equiv\left(c_{out}^{\dagger}+c_{out}\right)/\sqrt{2}$,
which contains the information of the spin operator $S_{z}^{2}$.
From Eq.$\,$(\ref{eq:Ry}), we obtain the measurement operator
\begin{equation}
M_{X}=\left(\pi\right)^{-\frac{1}{4}}\exp\left[i\eta S_{z}-\left(X_{L}+\chi_{x}S_{z}^{2}\right)^{2}/2\right]
\end{equation}
for measurement outcome $\left\langle \Upsilon\right\rangle =X_{L}=-\chi_{x}\left\langle S_{z}^{2}\right\rangle $,
which leads to the operation on the spin state
\begin{equation}
\left|\psi\right\rangle \mapsto M_{X}\left|\psi\right\rangle .
\end{equation}
To optimize the measurement on the amplitude (X) quadrature, we choose
a local pulse with $\beta_{LO}$ being proportional to $\beta_{2}$.
According to Eq.$\,$(\ref{eq:chix}), the spectrum of the optimal
input probe should satisfy $\beta_{in}^{2}\left(\omega\right)=8\kappa^{5}/\left[3\pi\left(\kappa^{2}+\omega^{2}\right)^{3}\right]$.
Therefore, we can obtain the optimal measurement strength $\chi_{x}=\sqrt{42N_{p}}\Omega^{2}/(2\kappa^{2})$
and the accumulated phase $\eta=-5\Omega N_{p}/\left(3\kappa\right)$
\cite{vanner2011selective} for X quadrature homodyne measurement.

Consider an atom ensemble initially prepared in a coherent spin state
(CSS) \cite{arecchi1972atomic} along the $x$ axis satisfying $S_{x}|\psi_{\mathrm{CSS}}\rangle=S|\psi_{\mathrm{CSS}}\rangle$,
which can be represented in the basis of the Dicke states as
\begin{equation}
|\psi_{\mathrm{CSS}}\rangle=\sum_{m=-S}^{S}N_{m}\left|S,m\right\rangle ,
\end{equation}
where $\left|S,m\right\rangle $ is the Dicke state with eigenvalue
of $m$ under the operator of $S_{z}$, $N_{m}=2^{-S}\binom{2S}{S+m}^{1/2}$
is the coefficients, and $S=N/2$ for the ensemble of $N$ atoms.
After the homodyne measurement, the final state of the collective
spins is given by
\begin{align}
\left|\psi_{f}\right\rangle  & =M_{X}|\psi_{\mathrm{CSS}}\rangle\nonumber \\
 & =\frac{\pi^{-\frac{1}{4}}}{\mathcal{N}}\sum_{m=-S}^{S}N_{m}e^{i\eta m-\frac{\left(X_{L}+\chi_{x}m^{2}\right)^{2}}{2}}\left|S,m\right\rangle ,\label{eq:MS}
\end{align}
where $\mathcal{N}$ is the normalization factor. Note that the state
after the measurement is a pure state and the action of measurement
operator is to prepare a superposition \cite{romero2011large} of
Dicke states.

\begin{figure}[htbp]
\center \includegraphics[width=1.05\columnwidth]{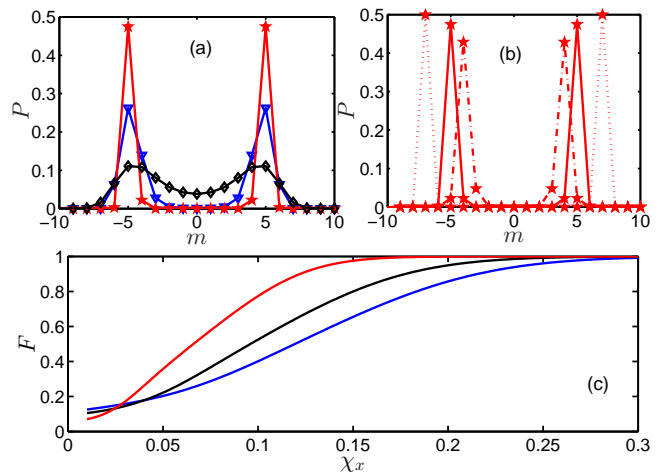}
\caption{(Color online) (a) and (b) Probability distribution of quantum superposition
states projected to Dicke states $\left|S,m\right\rangle $. The solid
lines show the probability distribution with $\chi_{x}=0.05$ (black),
$0.1$ (blue), and $0.2$ (red) for $X_{L}=-\chi_{x}S/2$ in (a) .
For the different measurement values, the probability is plotted in
(b) with $X_{L}=-\chi_{x}S/3$ (dashed), $-\chi_{x}S/2$ (solid),
and $-\chi_{x}S$ (dotted) for $\chi_{x}=0.2$. (c)The fidelity as
a function of $\chi_{x}$ for various measurement values : $X_{L}=-\chi_{x}S/3$
(blue), $-\chi_{x}S/2$ (black), $-\chi_{x}S$ (red). The system size
is $S=50$.}

\label{Fig2}
\end{figure}

From the Eq.$\,$(\ref{eq:MS}), we know that the collective spin
state collapses to a superposition of two wave-packets of Dicke states
concentrate around $m\approx\pm\sqrt{\frac{-X_{L}}{\chi_{x}}}$ after
the homodyne detection. The two wave-packets are separated by a distance
$d=2\sqrt{\frac{-X_{L}}{\chi_{x}}}$ along the direction of $S_{z}$,
where a width of the wave-packet is given by approximately $\sigma\sim\left(2\sqrt{-X_{L}\chi_{x}}\right)^{-1}$.
For an ideal measurement, where $m_{c}=\sqrt{\frac{-X_{L}}{\chi_{x}}}$
is an integer and $\left(\chi_{x}\pm\sigma^{-1}\right)^{2}\gg1$,
the quantum state is the superposition of two Dicke states as
\begin{align}
\left|\widetilde{\psi_{f}}\right\rangle  & \approx\frac{1}{\sqrt{2}}\left(e^{i\eta m_{c}}\left|S,m_{c}\right\rangle +e^{-i\eta m_{c}}\left|S,-m_{c}\right\rangle \right).\label{eq:MS1}
\end{align}

To intuitively illustrate the prepared quantum superposition states
by homodyne measurement, we calculate the probability distribution
$P\left(m\right)=\left|\left\langle S,m|\psi_{f}\right\rangle \right|^{2}$.
In Fig.$\,$\ref{Fig2}(a), we plot the $P\left(m\right)$ as a function
of $m$ with different $\chi_{x}$, where we choose the outcome of
measurement $X_{L}=-\chi_{x}S/2$. When the measurement strength $\chi_{x}$
is weak, the state is a superposition of many Dicke states, and we
can observe two wave packets with central position $m\approx\pm\sqrt{\frac{S}{2}}$.
With the increasing $\chi_{x}$, the width of the wavepackets becomes
narrower while the peaks become higher. When $\chi_{x}=0.2$, the
state almost consists of only two Dicke states, which is separated
by a distance $d=2\sqrt{\frac{S}{2}}$. For different measurement
outcomes $X_{L}$, we show the probability distribution in Fig.$\,$\ref{Fig2}(b).
If $X_{L}$ becomes larger, the two wave-packets are further apart
and the width of the wave-packets becomes narrower, which is consistent
with the theoretical analysis. When the parameters satisfy the condition
$\left(\chi_{x}\pm\sigma^{-1}\right)^{2}\gg1$, the state is only
the superposition of two Dicke states. In Fig.$\,$\ref{Fig2}(c),
we select the superposition of two Dicke states from Eq.$\,$(\ref{eq:MS1})
as the target state and calculate the fidelity $F=\left|\left\langle \widetilde{\psi_{f}}|\psi_{f}\right\rangle \right|^{2}$
as a function of $\chi_{x}$. We find that the fidelity $F$ approaches
$1$ with the increasing $\chi_{x}$. For different measurement outcomes,
the required $\chi_{x}$ for near-unit fidelity state preparation
is different, and the $\chi_{x}$ becomes larger with the larger measurement
outcomes $X_{L}$. Specifically, we can also realize the GHZ state
$\left|\widetilde{\psi_{f}}\right\rangle \sim e^{i\eta S}\left|S,S\right\rangle +e^{-i\eta S}\left|S,-S\right\rangle $
when $X_{L}=-\chi_{x}S^{2}$, which is particularly important for
quantum information processing.

\section{Dicke squeezed states}

When choosing $\phi=\pi/2$, we measure the phase (P) quadrature $\Upsilon^{out}\left(\pi/2\right)=P^{out}\equiv i\left(c_{out}^{\dagger}-c_{out}\right)/\sqrt{2}$.
Similar to the measurement on the amplitude quadrature, the phase
quadrature $P^{out}$ of output light contains information about the
spin operator $S_{z}$, and leads to the actions on the spin operator
after measurement. From Eq.$\,$(\ref{eq:Ry}), the measurement operator
\cite{vanner2011selective,vanner2011pulsed} with $\left\langle \Upsilon\right\rangle =P_{L}=-\chi_{p}\left\langle S_{z}\right\rangle $
is described as
\begin{equation}
M_{P}=\pi^{-\frac{1}{4}}\exp\left[i\eta S_{z}-\left(P_{L}+\chi_{p}S_{z}\right)^{2}/2\right],
\end{equation}
and the output state
\begin{equation}
\left|\psi\right\rangle \mapsto M_{P}\left|\psi\right\rangle .
\end{equation}
For an optimal measurement of the spin operator \cite{vanner2011selective},
the local input pulse $\beta_{LO}$ is chosen to have an amplitude
directly proportional to $\beta_{1}$. When $\beta_{in}(t)=\sqrt{\kappa}e^{-\kappa\left|t\right|}$
\cite{vanner2011pulsed}, we have the optimal measurement strength
$\chi_{p}=\sqrt{10N_{p}}\Omega/\kappa$ and accumulated phase $\eta=-\frac{3\Omega N_{p}}{2\kappa}$.

Similar to the preparation of superposition state, we start with an
atomic system prepared initially in a CSS along the $x$ axis, and
the output state after measurement is
\begin{align}
\left|\psi_{f}\right\rangle  & =M_{P}\left|\psi_{\mathrm{CSS}}\right\rangle \nonumber \\
 & =\frac{\pi^{-\frac{1}{4}}}{\mathcal{N}}\sum_{m=-S}^{S}N_{m}e^{i\eta m-\frac{\left(P_{L}+\chi_{p}m\right)^{2}}{2}}\left|S,m\right\rangle .
\end{align}
The equation indicates that the state has one wave-packet of Dicke
states, where the center of the wave-packet is $m=-P_{L}/\chi_{p}$.
If $\chi_{p}^{2}\gg1$, we can obtain the Dicke state as $\left|\psi_{f}\right\rangle =\left|S,m\right\rangle $
after one-shot measurement, where $m$ is an integer around the $-P_{L}/\chi_{p}$.

Specially, if $P_{L}=0$, we obtain a Dicke state $\left|S,0\right\rangle $
which is known as the DSS that is useful for quantum metrology \cite{opatrny2016counterdiabatic}.
To characterize the DSSs, we introduce a single experimentally detectable
parameter as the figure of merit in quantum metrology \cite{zhang2014quantum}
\begin{equation}
\xi_{D}=\frac{N\left(\left\langle \left(\Delta S_{z}\right)^{2}\right\rangle +1/4\right)}{\langle S_{x}^{2}+S_{y}^{2}\rangle}.
\end{equation}
It is worth noting that the squeezing parameters $\xi_{D}$ is universal
for characterizing the entanglement depth of all DSSs \cite{vitagliano2017entanglement}.
In addition, the DSSs described by $\xi_{D}$ are more robust to decoherence
and experimental noise than other quantum states \cite{zhang2014quantum}.
For a CSS along the $x$ axis, we have $\xi_{D}=1$, thus a state
is the Dicke spin-squeezed state when $\xi_{D}<1$. The parameter
$\xi_{D}$ can attain the minimum $1/\left(N+2\right)$ under the
ideal DSS $\left|S,0\right\rangle $, so the phase sensitivity of
DSS approaches the Heisenberg limit.

\begin{figure}[htbp]
\center \includegraphics[width=0.95\columnwidth]{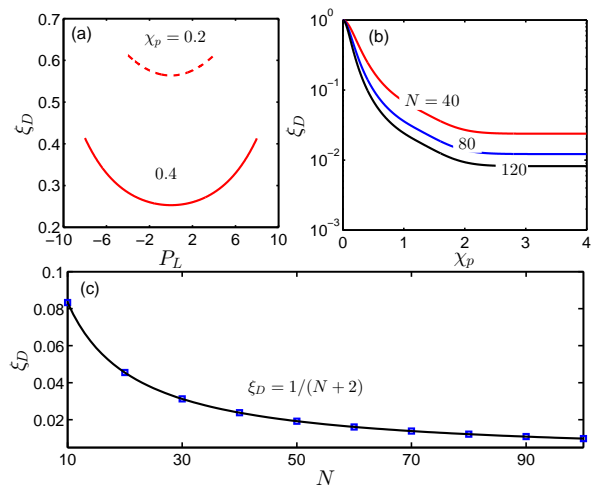}
\caption{(Color online) (a) The squeezing parameter $\xi_{D}$ as a function
of the phase-quadrature measurement $P_{L}$ for $N=40$ and $\chi_{p}=0.2,~0.4$.
(b) The squeezing parameter as a function of the spin-measurement
strength $\chi_{p}$ for $P_{L}=0$ and $N=40,~80,~120$. (c) Optimal
spin squeezing plotted against the number of atoms $N$ with the analysis
result (solid line) and numerical result (square dots).}

\label{Fig3}
\end{figure}

In Fig.$\,$\ref{Fig3}(a), we plot the spin squeezing parameter $\xi_{D}$
as a function of the phase-quadrature measurement $P_{L}$ for both
$\chi_{p}=0.2$ (dotted red line) and $\chi_{p}=0.4$ (solid red line),
where $P_{L}\in\left[-\chi_{p}S,~\chi_{p}S\right]$. It clearly shows
that spin squeezing parameter $\xi_{D}$ reaches its minimal when
$P_{L}=0$, and the minimal value decreases with the increasing of
$\chi_{p}$. The Fig.$\,$\ref{Fig3}(b) further shows that the squeezing
parameter increases with the measurement strength $\chi_{p}$, and
the $\xi_{D}$ saturates at certain value that reduces with atom numbers
$N$. The reason for the saturation behavior of $\xi_{D}$ is that
the larger measurement strength carries more information about the
spin operator, which leads better squeezing. When $\chi_{p}\geq2$,
we can obtain the ideal DSS $\left|S,0\right\rangle $. To further
verify that optimal spin squeezing is what we expect, the optimal
squeezing as a function of $N$ is shown in Fig.$\,$\ref{Fig3}(c).
The black line is the analysis result $\xi_{D}=1/\left(N+2\right)$
for the ideal DSS, and all square dots (numerical result) are on the
black line. These demonstrate that theoretical analysis and numerical
results are consistent with each other, and we can realize the DSSs
that approaching the Heisenberg limit for precision measurement \cite{zhang2014quantum}.

\section{Discussion}

For the experimental realization of our scheme, we consider a dilute
ensemble of $^{87}\mathrm{Rb}$ trapped inside an optical cavity \cite{leroux2010implementation,schleier2010squeezing},
with $g=2\pi\times0.4~\mathrm{MHz}$, $\Delta=2\pi\times3~\mathrm{GHz}$
and $\kappa=2\pi\times1~\mathrm{MHz}$. In our model, the approximation
of adiabatical elimination of the excited state of atoms \cite{zhang2015detuning}
is valid only when the intracavity photon number be sufficiently low,
which should satisfy $\left\langle c^{\dagger}c\right\rangle \ll(\Delta/g)^{2}$.
From Eq.$\,$(\ref{eqC-1}), we can obtain the intracavity photon
number$\left\langle c^{\dagger}c\right\rangle \simeq N_{p}\left|\beta_{0}(t)\right|^{2}$.
For homodyning the amplitude quadrature using the optimal input probe,
we obtain the intracavity $\left\langle c^{\dagger}c\right\rangle \leq4N_{p}$.
Therefore, the amplitude measurement strength satisfies $\chi_{x}\ll\sqrt{42}g^{3}/\left(\kappa^{2}\Delta\right)\approx1.4\times10^{-4}$,
which is very weak for low intracavity photons. For homodyning the
phase quadrature, we have the intracavity photon number $\left\langle c^{\dagger}c\right\rangle \leq2N_{p}/e$,
and the phase measurement strength is $\chi_{p}\ll\frac{g\sqrt{20e}}{\kappa}\approx3$.
It means that we can not use the very strong probe pulse.

\begin{figure}[htbp]
\center \includegraphics[width=0.95\columnwidth]{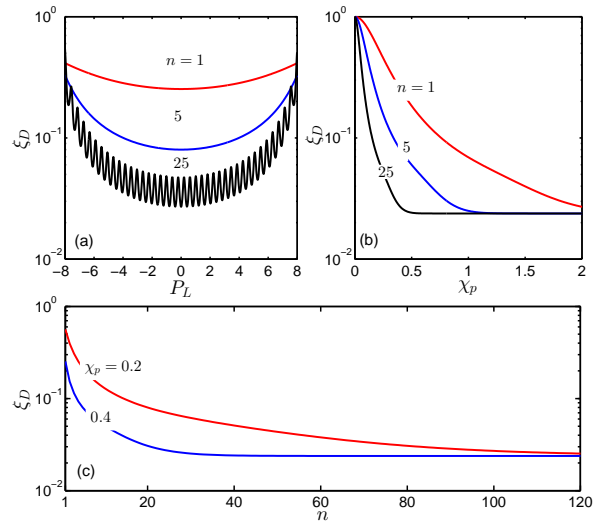}
\caption{(Color online) (a) Spin squeezing parameter $\xi_{D}$ as a function
of the phase-quadrature measurement $P_{L}$ for $\chi_{p}=0.4$ and
$n=1,~5,~25$. (b) Spin squeezing parameter as a function of the spin-measurement
strength $\chi_{p}$ for $P_{L}=0$ and $n=1,~5,~25$. (c) Spin squeezing
plotted against the number of measurement times $n$ for $P_{L}=0$
and $\chi_{p}=0.2,~0.4$. (a)-(c) $N=40$.}

\label{Fig4}
\end{figure}

We propose to enhance the fidelity of state preparation by repetitive
measurements $\prod_{j=1}^{n}M^{j}$ of the cavity-atom ensemble system,
where $n$ is the rounds of measurements. From the effective projection
operator
\begin{align}
\frac{M^{eff}}{\left(\pi\right)^{-\frac{n}{4}}} & =\prod_{j=1}^{n}\exp\left[i\eta^{j}S_{z}-\frac{\left(\Upsilon^{j}+\chi_{x}^{j}S_{z}^{2}+\chi_{p}^{j}S_{z}\right)^{2}}{2}\right],
\end{align}
the effective measurement strength could be enhanced collectively
by a factor of $\sqrt{n}$. For example, by simply assuming that every
measurement outcome is the same in preparing quantum superposition
of Dickes states, we obtain the effective outcome of the output $X_{L}^{eff}=\sqrt{n}X_{L}$
and effective measurement strength $\chi_{x}^{eff}=\sqrt{n}\chi_{x}$.
Similarly, the preparation of DSS can also be enhanced by the repetitive
measurements . By assuming that both the input pulses and measurement
results remain unchanged with both $P_{L}^{j}=0$ and $\chi_{p}^{j}=\chi_{p}$,
then we can obtain the effective measurement strength $\chi_{p}^{eff}=\sqrt{n}\chi_{p}$.
In Fig.$\,$\ref{Fig4}, we investigate the spin squeezing by repetitive
measurement, and the complex effects of $n$ on the squeezing parameters
$\xi_{D}$ is observed. For weak measurement-strength $\chi_{p}=0.4$,
the degree of squeezing $\xi_{D}$ increase with $n$. For $n=25$,
the spin squeezing parameter has a small fluctuation in the range
$P_{L}\subseteq\left[-\chi_{p}S,~\chi_{p}S\right]$, which means that
the spin squeezing can be enhanced for a wide range of the measured
value $P_{L}$. The dependence of spin squeezing on $\chi_{p}$ is
presented in Fig.$\,$\ref{Fig4}(b) for $n=1,~5,~25$. According
to the $\sqrt{n}$-enhancement, the optimal spin squeezing should
be obtained when $n\approx\left(\chi_{p}^{\mathrm{opt}}/\chi_{p}\right)^{2}=\left(2/\chi_{p}\right)^{2}$,
which agrees well with the Fig.$\,$\ref{Fig4}(c).

For the measurement-based quantum state preparation, the drawback
is that the desired processes depends on the measurement outcome and
is probabilistic. For the repetitive measurement enhanced spin squeezing
state preparation, $n=100$ ($50$) measurements are required to achieve
optimal spin squeezing for $X_{p}=0.2$ ($0.4$). The probability
of the measurement $P_{L}=0$ exponentially decreases with $n$, thus
is very challenging for experimental realizations. Actually, we find
that the sufficiently low intracavity photon number $\left\langle c^{\dagger}\left(t\right)c\left(t\right)\right\rangle \ll(\Delta/g)^{2}$
is an instantaneous condition, so we use the long pulse probing with
$\beta_{in}(t)=\sqrt{\kappa/n_{t}}e^{-\kappa\left|t\right|/n_{t}}$
to reduce the instantaneous intracavity field, where $n_{t}/\kappa$
is the effective length of the pulse. Therefore, we can realize the
optimal spin squeezing by combining long pulse probing and repetitive
measurements.

We can obtain the $\left\langle c^{\dagger}\left(t\right)c\left(t\right)\right\rangle \leq2N_{p}/\left(\sqrt{n_{t}}e\right)$,
which means $\chi_{p}\ll\frac{g\sqrt{20n_{t}e}}{\kappa}\approx3\sqrt{n_{t}}$.
We can realize the optimal spin squeezing with both the long pulse
and strong probing. It is best to realize spin squeezing in experiment
by combining long pulse probing and multiple measurements. If we choose
$n_{t}=10$, the probability of optimal spin squeezing can be obtained
greatly increased only by $n=4$ times of the homodyne detection.

\section{Conclusion}

We propose an experimental feasible scheme to prepare the quantum
states of an atomic ensemble via cavity-assisted homodyne measurements.
It is revealed that the probe light that resonant with the cavity
would induce two distinct effects: one is that quantum sate of output
light can be entangled with that of the atomic ensemble in the cavity,
the other one is the light mediated indirect nonlinear interaction
between atoms. By selectively measuring the quadrature of the output
light, we propose the preparation of the superposition of Dicke states
and the DSS. The scheme is feasible for experiments, as the quantum
state preparation with weak measurement strength can be enhanced by
repetitive homodyne detection or using longer probing laser pulses.
Our scheme also holds the potential to generate other non-Gaussian
quantum states of atom ensemble, which may find applications in future
studies of quantum mechanics and quantum metrology.

{\em Acknowledgments.} This work was funded by the National Key
R \& D Program (Grants No. 2016YFA0301300) and the National Natural
Science Foundation of China (Grants No. 11474271, No. 11674305, No.
11704370 and No. 61505195), and Anhui Initiative in Quantum Information
Technologies (AHY130000). T. X. is supported by the National Natural
Science Foundation of China (Grants No. 91636215 and No. 117043658)
and Chinese Academy of Science (Grant No. XDB21010200).

\bibliographystyle{Zou}

\end{document}